# The Quantum Capacitor Detector: A Single Cooper Pair Box Based Readout for Pair Breaking Photo-detectors

P.M. Echternach, M.D. Shaw[*], J. Bueno, P.K.Day, C.M. Bradford
*Jet Propulsion Laboratory, California Institute of Technology, Pasadena CA 91109, USA*
[*]*University of Southern California, Department of Physics and Astronomy, Los Angeles CA 90089-0484, USA*

We propose a sensitive new detector based on Cooper pair breaking in a superconductor. The quantum capacitor detector (QCD) exploits the extraordinary sensitivity of superconducting single-electron devices to the presence of quasiparticles generated by pair-breaking photons. This concept would enable single-photon detection at far-IR and sub-millimeter frequencies with detector sensitivities that exceed that of transition-edge-sensor bolometers (TES), kinetic inductance detectors (KID), and superconducting tunnel junction detectors (STJ). The detectors we propose are based on the single Cooper pair box (SCB), a mesoscopic superconducting device that has been successfully developed at JPL for applications in quantum computing. This concept allows for frequency multiplexing of a large number of pixels using a single RF line, and does not require individual bias of each pixel. The QCD is ideal for the sensitive spectrographs considered for upcoming cold space telescopes, such as BLISS for SPICA in the coming decade, and for the more ambitious instruments for the SAFIR / CALISTO and SPIRIT / SPECS missions envisioned for the 2020 decade. These missions require large detector arrays (> 10,000 elements) which are limited by astrophysical background noise, corresponding to a noise-equivalent power (NEP) as low as $2 \times 10^{-20}$ W / Hz$^{1/2}$. Given its intrinsic response time, the QCD could also be used for energy-resolved visible photon detection, with estimated E / $\Delta$E > 100, enabling imaging low-resolution spectroscopy with an array of detectors.

The detector we propose is an example of a superconducting pair-breaking detector. Photons with energy hν > 2Δ, where Δ is the superconducting gap, can break Cooper pairs (weakly bound pairs of electrons which are charge carriers in the superconducting state) generating quasiparticles, which can be thought of as unpaired free electrons. The problem faced by a pair-breaking detector is how to measure quasiparticles in the presence of the Cooper pairs. For example, in a KID [1] the quasiparticle population changes the inductance and center frequency of a superconducting resonator. In an STJ detector [2] the quasiparticles are filtered out of the Cooper-pair background and measured as a current though a tunnel junction. STJ detectors can deliver NEPs on the order of $10^{-19}$ W/Hz$^{1/2}$ but suffer from a few disadvantages when scaled to large arrays: (i) a magnetic field is applied to suppress the Josephson (Cooper pair) current, but junction size variations make it difficult for the same value of the field to suppress that current from all junctions; (ii) multiplexing is proposed using the RF single-electron transistor (RF-SET) [3, 4], but the number of pixels would be limited, since impedance matching requirements dictate a resonator Q of the order 30, so each pixel would occupy a wide frequency band. KIDs can deliver multiplexing of a large number of pixels since the resonator Qs are large ($10^5$ to $10^6$). However, their performance is currently limited by resonator phase noise caused by two-level charge fluctuators, and present NEP values are on the order of $10^{-18}$ W/Hz$^{1/2}$.

The QCD is inherently faster (~100μs) than a TES bolometer [5] since the response time of the TES is limited by the very small thermal conductance G needed to achieve low NEP. The response time of an ultra-sensitive TES can be of the order of 1 s. In addition, TES arrays require time domain multiplexing, which has not been experimentally demonstrated in devices with ultra-low NEP.

Our proposed device makes use of the single Cooper pair box (SCB) [6-8], a mesoscopic superconducting device that has been successfully developed at JPL for applications in quantum computing. The SCB consists of a small superconducting island (the "box") coupled to a larger volume of superconducting material known as the "reservoir". The coupling is made via a pair of ultra-small tunnel junctions, so that single Cooper pairs can tunnel onto the island individually. An electrode fabricated in close proximity to the island forms a gate capacitor which can be used to adjust the island potential (see Figure 1). At a particular value of gate charge known as the "degeneracy point", the energy levels of the system form an isolated two- state quantum system. To measure the state of the system, we measure the "quantum capacitance" of the device by placing it in a parallel LC tank circuit, which is monitored with RF reflectometry. If quasiparticles are present in the reservoir, they can also tunnel onto the island, effectively shifting the gate charge by one electron and changing the overall capacitance by a few fF [9-12]. In a typical experiment, a single quasiparticle tunneling across the junction will produce a reflected signal phase shift of approximately 140º (see Figure 2). Since the devices are operated at temperatures much lower than the superconducting gap energy, background signal from quasiparticles in equilibrium will be exponentially suppressed. Figure 1 shows two coupled SCBs with multiplexed on-chip LC tank circuits fabricated at JPL for research in quantum computing. We have recently performed extensive measurements and modeling of quasiparticle tunneling in such devices [12]. Figure 2A shows real time measurements of the phase of the reflected signal at the degeneracy point. The jumps correspond to the tunneling of a single quasiparticle onto or off of the island. The data can be plotted in a histogram, as shown in Figure 2B, from which the phase noise of the measurements can be extracted. By plotting the histograms as a function of gate voltage, such as the graph shown in Figure 2C, we can trace the expected phase shift of the reflected signal for an SCB with and without an extra quasiparticle in the island. In our concept, as shown in Figure 1, radiation is coupled via an antenna to a small superconducting reservoir, where the absorbed energy generates quasiparticles by breaking Cooper pairs. The reservoir has a lower superconducting gap than that of the antenna, so quasiparticles generated in the antenna are concentrated in the reservoir. For a photon with energy $h\nu$ and a reservoir material with superconducting gap $\Delta$, a number of quasiparticles $N_{qp}=\eta\, h\nu\,/\Delta$ is generated. Here $\eta \approx 0.57$ is the efficiency for conversion of the photon energy into quasiparticles [13]. With the SCB biased at the degeneracy point, the box will behave as a potential well or "trap" for the quasiparticles with a depth $\delta E \sim E_C - E_J/2$, where $E_c$ is the charging energy given by $e^2/2C_\Sigma$, $E_J$ is the Josephson Energy and $C_\Sigma$ is the total capacitance of the island, including the two tunnel junctions and the gate capacitance. Quasiparticles tunnel onto the island with a tunneling rate $\Gamma_{eo}(n_{qp})$ which depends linearly on the density of quasiparticles in the reservoir $n_{qp}=N_{qp}/\Omega_L$, where $\Omega_L$ is the reservoir volume. Quasiparticles will tunnel off the island with a rate $\Gamma_{oe}$ which is approximately independent of the number of quasiparticles in the reservoir. At steady state, the probability of a quasiparticle being present in the island is given by $P_o(N_{qp})=\Gamma_{eo}/(\Gamma_{eo}+\Gamma_{oe})$ and the resulting change in the average capacitance will be $C_Q= (4E_C/E_J)(C_g^2/C_\Sigma)P_o(N_{qp})$. With the tank circuit tuned to its resonance frequency $\omega_o$, this change in capacitance will produce a phase shift $\delta\Phi \sim 2C_Q/(\omega_o Z_o C_C^2)$ in the reflected signal, where $Z_o$ is the characteristic impedance of the RF line (nominally 50Ω) and $C_C$ is the coupling capacitance to the tank circuit. Using measured values of $\Gamma_{oe}$ and $\Gamma_{eo}$ [12] we estimate the phase shift at 18mK due to a single quasiparticle tunneling event to be approximately 2.4 radians or 138 degrees, which is close to the measured values such as those shown in Figure 2. The responsivity of the detector, the change in the phase of the reflected signal as a function of the number of quasiparticles, is given by

$$\frac{d\Phi}{dN_{qp}} = \frac{2}{Z_o} \frac{1}{\omega_o C_C^2} \frac{C_g^2}{C_\Sigma} \frac{4E_C}{E_J} \frac{\Gamma_{eo}(N_{qp})\Gamma_{oe}/N_{qp}}{(\Gamma_{eo}(N_{qp})+\Gamma_{oe})^2}.$$

*Noise Sources*
  *a) Phase Noise*
  The quantum capacitance of the SCB has been measured routinely in our laboratory, with a resonant circuit with the same parameters as the ones proposed for QCD readout. Figure 2B shows a typical phase shift histogram, the width of which gives the phase noise of our measurement. This phase noise includes amplifier noise and background charge fluctuation noise. Fitting this histogram to a Gaussian, we find an rms noise of 33 degrees, and considering the measurement bandwidth of 100kHz, we estimate a phase noise of 0.1 degrees/Hz$^{1/2}$ or $\delta\Phi \sim 1.8 \times 10^{-3}$ radian/Hz$^{1/2}$. Therefore, the minimum number of quasiparticles we can detect is $\delta N_{qp} \sim 10^{-3}$ qp/Hz$^{1/2}$. The NEP associated with the phase measurement noise can then be estimated by $NEP_\Phi = (\Delta/\eta\, \tau_R)\, \delta N_{qp}$. Assuming a recombination time of about 100μs (measured by Day et. al in similar Al films [1]) and the superconducting gap of Aluminum (2.1 K) we estimate an NEP due to phase noise $NEP_{PN} \sim 5 \times 10^{-21}$ W/Hz$^{1/2}$. As can be seen in Figure 3, phase noise is not the dominant source of noise in QCD devices.
  *b) Shot noise*

The quasiparticle tunneling process from the lead to the island is a Poisson process, so the tunnel rate has intrinsic fluctuations $\delta\Gamma_{eo} = \sqrt{\Gamma_{eo}}$, and $\delta\Gamma_{oe} = \sqrt{\Gamma_{oe}}$, and a corresponding fluctuation in the number of quasiparticles in the reservoir $\delta N_{qp}$, which will in turn cause fluctuations in the signal phase shift. This is indicated by the green curve in Figure 3A. Quasiparticle shot noise is the leading source of noise at low temperatures.

*c) Generation recombination noise*

Quasiparticles can be thermally excited over the superconducting gap and recombine into Cooper pairs, introducing a fluctuation in the number of quasiparticles in the reservoir known as the generation-recombination noise [14]. This is indicated by the red curve in figure 3A. Note that generation-recombination noise is the dominant noise source at high temperatures, where equilibrium quasiparticle states begin to be populated.

*Noise equivalent power*

From our measurements of the tunneling rates $\Gamma_{eo}$ and $\Gamma_{eo}$ [12] and the theoretical model of Lutchyn and Glazman [15] we were able to calculate the contributions of all noise sources to the noise-equivalent power. Figure 4 shows the estimated NEP from the various noise sources described above, using the parameters of a device optimized for BLISS. From the results, we see that the devices can be operated at any temperature below 100mK, at which the generation-recombination noise starts to dominate. Choosing an operating temperature of 80mK, we calculated the performance for the range of wavelengths appropriate for BLISS, as shown in figure 3B. One interesting observation is that the reservoir size in this design is comparable to the ground plane volume of our quantum computing samples. In essence, SCB qubits are already operating as detectors of electromagnetic noise with NEP in the range of $10^{-19}$W/Hz$^{1/2}$.

*Tunneling time*

Once quasiparticles are generated in the reservoir, they must on average tunnel onto to the SCB island in a time shorter than the recombination time. To tunnel, quasiparticles generated in the reservoir must move diffusively to the junctions. The average time it will take a quasiparticle to sample the entire volume of the reservoir is given by $\tau = A/D$ where $A$ is the reservoir area and $D$ is the diffusion constant. For $A=2.5 \times 10^{-6} cm^2$ (corresponding to our reservoir design volume of $5 \times 10^{-18} m^3$ with a 20nm thick film) and a diffusion constant $D = 35 cm^2/s$ (measured in actual samples in our laboratory) this time would be $\tau = 71$ ns, three orders of magnitude shorter than the quasiparticle recombination time ($\tau_R \sim 100 \mu s$).

We now evaluate the tunneling time, the time it takes one quasiparticle to tunnel through the junctions once it is generated in the reservoir. To estimate the tunneling time, we use our model for the rate $\Gamma_{eo}$ [12], and take the tunneling time to be $\tau_t = 1/\Gamma_{eo}(1)$, where $\Gamma_{eo}(1)$ is the even to odd tunneling assuming one quasiparticle present in the reservoir. Making that assumption in our model, we obtain $\tau_t \approx 20 \mu s$, a time five times smaller than the recombination time. Note that this estimated tunneling time is much larger than the time calculated for classical devices with standard formulas available in the literature [16] which yield $\tau_t \approx 0.08 \mu s$. We are confident that our model captures the essential features of the physical system and provides a more reliable estimate. Regardless, the tunneling time is still short compared to the quasiparticle recombination time.

*Detector speed*

Since the quasiparticles are trapped in the reservoir after being generated, the speed of the detector will be ultimately limited by the quasiparticle recombination time $\tau_R \approx 100 \mu s$.

*Dynamic Range*

Figure 4 shows the NEP of the QCD detector as a function of loading power at a 30μm wavelength. Also shown is the photon noise as a function of power at 30 μm. The detector saturation power will be defined as the power at which the detector NEP becomes equal to the photon noise. The dynamic range is the ratio of the saturation power over the BLISS loading power at that wavelength. At 30 μm, the saturation power is $3 \times 10^{-15}$W, and the dynamic range 18552. Note that the detector will still work above the saturation power, but in this case it will add noise of the photon noise. The table summarizes the performance of the QCD detector over the range of BLISS.

| Wavelenght (μm) | Loading ($10^{-19}$W) | Required NEP ($10^{-20}$W/Hz$^{1/2}$) | QCD NEP ($10^{-21}$W/Hz$^{1/2}$) | Saturation power ($10^{-17}$W) | Dynamic Range |
|---|---|---|---|---|---|
| 30 | 1.62 | 4.51 | 3.13 | 300 | 18552 |
| 45 | 1.46 | 3.51 | 2.96 | 200 | 13661 |
| 70 | 1.34 | 2.69 | 2.83 | 100 | 7477 |
| 100 | 1.81 | 2.62 | 3.32 | 60.0 | 3314 |
| 140 | 2.94 | 2.81 | 4.37 | 43 | 1463 |
| 200 | 3.78 | 2.67 | 5.07 | 25.4 | 674 |
| 250 | 3.64 | 2.35 | 4.96 | 24.2 | 667 |
| 300 | 4.97 | 2.53 | 5.97 | 15.0 | 302 |
| 400 | 19.5 | 4.38 | 14.9 | 7.83 | 40 |
| 600 | 165 | 10.36 | 78.4 | 5.22 | 3 |
| 800 | 540 | 16.13 | 204 | 2.63 | 0.5 |

In conclusion, we proposed here a new type of Pair-Breaking detector with readout based on the Quantum Capacitor of a Single Cooper Pair Box. We present measurements of quasiparticle tunneling rates on SCBs that provide input to our model predicting the performance of the detector. We predict this detector will fulfill the stringent requirements of future Far Infrared missions using cold telescopes in space.

We would like to thank Richard Muller for performing the electron-beam lithography. This work was conducted at the Jet Propulsion Laboratory, California Institute of Technology, under a contract with the National Aeronautics and Space Administration, and was partially funded by a grant from the National Security Agency. Juan Bueno is supported by the NASA Postdoctoral Program. Copyright 2008, California Institute of Technology. Government sponsorship acknowledged.

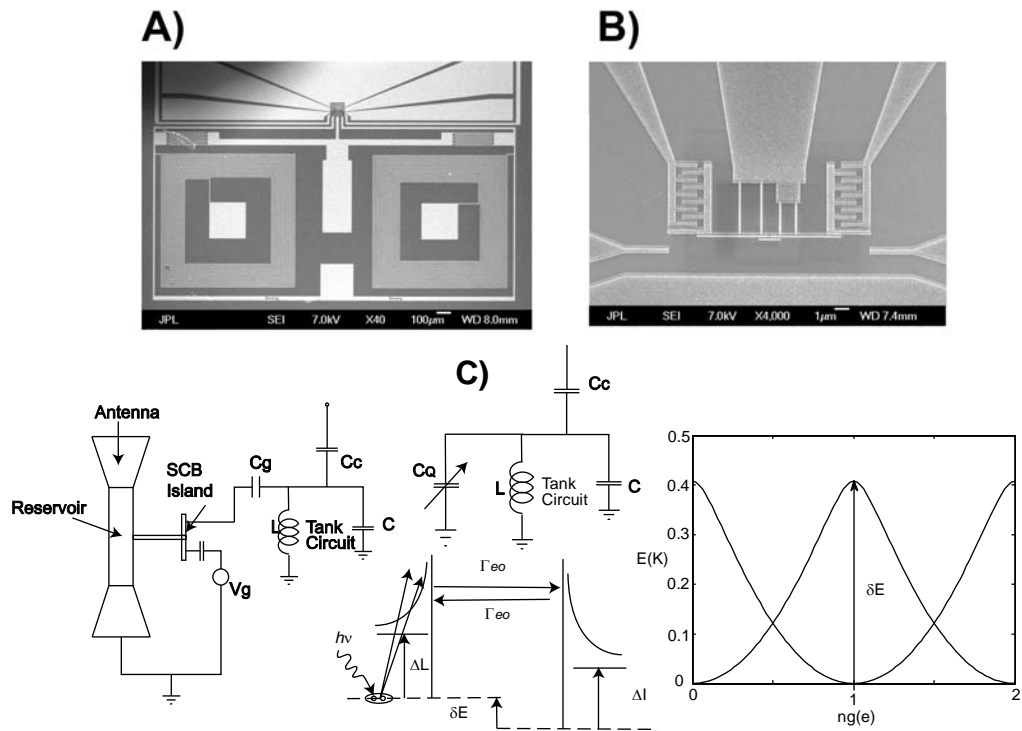

Figure 1. A) Scanning electron micrograph of multiplexed on-chip LC oscillators. B) Scanning electron micrograph of two side by side SCB structures. C) Concept of Quantum Capacitor Detector. $N_{qp}$ quasiparticles are created by breaking Cooper Pairs in a reservoir by antenna coupled radiation. Quasiparticles will tunnel to the island of the SCB which acts like a potential well of depth $\delta E$ with a rate $\Gamma_{eo}(N_{qp})$ and out of the well with a rate $\Gamma_{oe}$ which is not a function of $N_{qp}$. The quantum capacitance of the island changes by a value proportional to the quantity $P_e = \Gamma_{oe}/(\Gamma_{oe}+\Gamma_{eo})$ and shifts the tank circuit resonance frequency and consequently the phase of an RF signal reflected off the tank circuit.

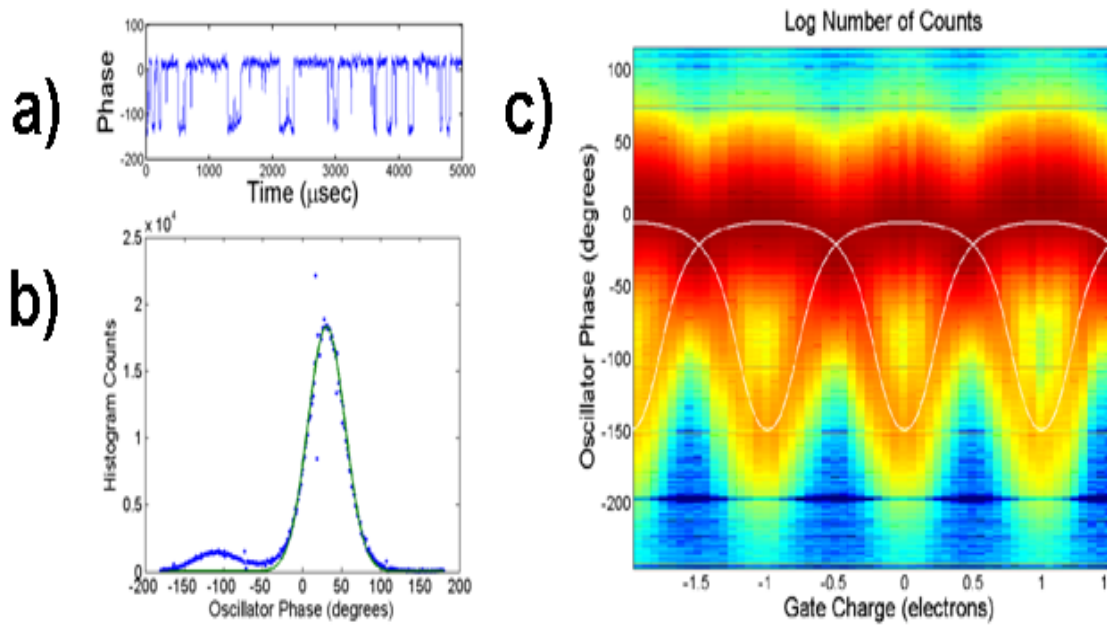

Figure 2. Measurements of the phase of the reflected signal from an SCB embedded in an LC tank circuit. A) Measurement of the phase as a function of time at the degeneracy point. The jumps correspond to a single quasiparticle tunneling on and off the island. Note the large phase shift (140 degrees) a quasiparticle can cause. B) Histogram of phase measurements showing an rms phase noise of 33 degrees or $1.8 \times 10^{-3}$ radians/Hz$^{1/2}$ given a measurement bandwidth of 100 kHz C) Plot of phase histograms as a function of gate voltage. The position of the peaks trace the phase as a function of gate voltage for an SCB with and without an extra quasiparticle. White curves are theoretical estimates of quantum capacitance.

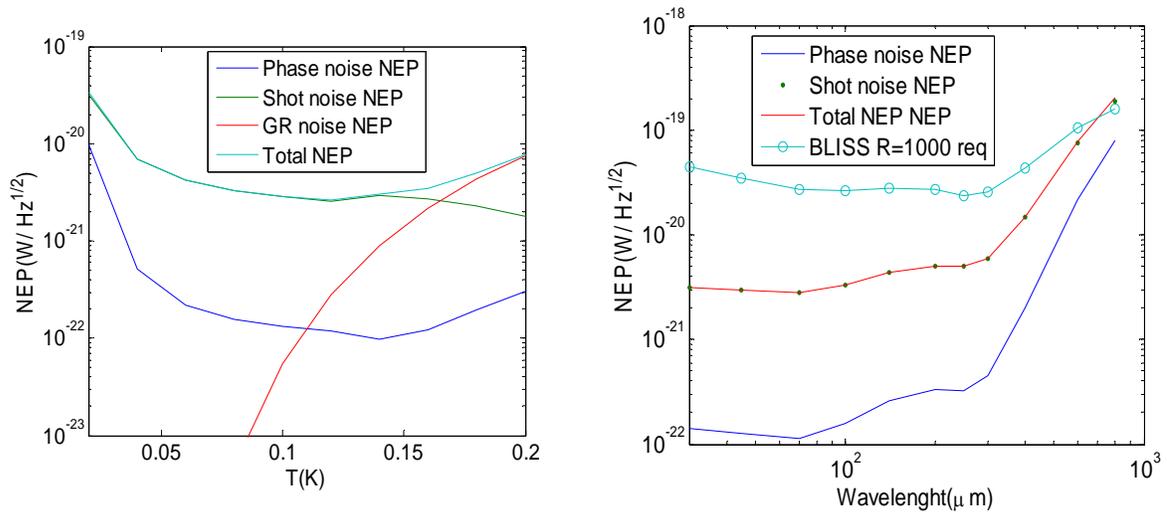

Figure 3. Left: NEPs from various noise sources calculated for devices optimized for BLISS (with 100μm loading, R=1000) as a function of temperature. Right: NEPs of various noise sources as a function of wavelenght as compared to the requirements for a BLISS type spectrometer with R=1000 . The operating temperature was chosen to be 0.08K.

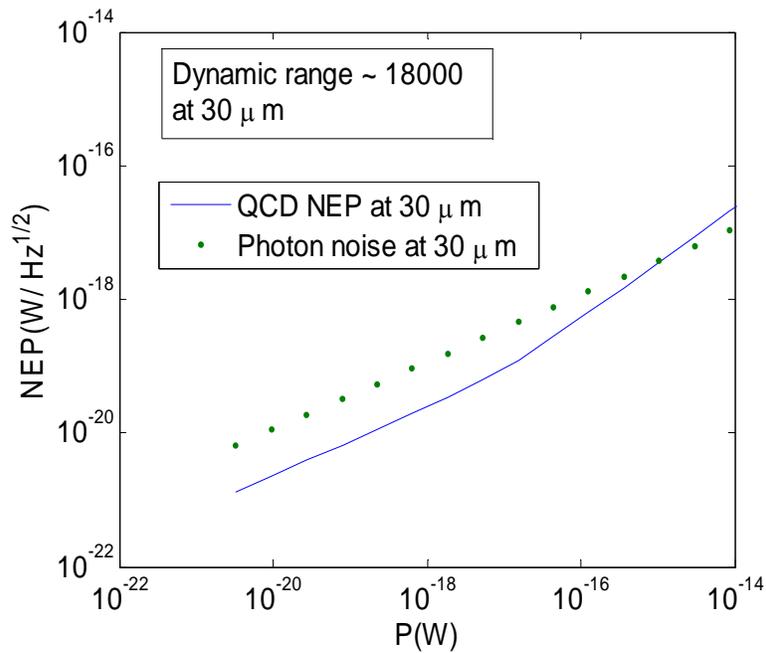

Figure 4. NEP of QCD detector as a function of loading power at 30μm as compared with the photon noise at the same wavelength. We define the saturation as the power at which the NEP becomes equal to the photon noise. The dynamic range is the ratio between the saturation power and the BLISS loading at that wavelength. The dynamic range at 30 μm is over 18000. Note that the detector still works for powers above the saturation, except that it will add noise over the photon noise.